\documentclass[10pt,preprint]{aastex}
\usepackage{psfig}

\def\eq#1{equation (\ref{#1})}

\def\au{{\rm AU}} 

\def\gaia{{\it Gaia}} 
\def\deg{^\circ} 
\def\msun{M_\odot}
\def\mearth{M_\oplus} 
\def\mjup{M_{\rm jup}} 
\def\muas{\mu{\rm as}}
\def\mx{M_X} 
\def\dx{D_X} 
\def\db{D_b} 
\def\pix{\Pi_X} 
\def\pis{\Pi_S}
\def\mux{\mu_X} 
\def\pirel{\Pi_{rel}} 
\def\be{\begin{equation}}
\def\ee{\end{equation}} 
\def\thetae{\theta_{\rm E}} 
\def\bth{{\bf \theta}}
\def\bu{{\bf u}} 
\def\siga{\sigma_{ast}} 
\def\sn{{\rm S/N}}
\def\sntot{(\sn)_{\rm tot}} 
\def\ave#1{\langle#1\rangle} 
\def\fsky{f_{\rm sky}}

\begin{document}

\title{Astrometric Microlensing Constraints on a Massive Body in the Outer
Solar System with \gaia} 
\author {B.\ Scott Gaudi\altaffilmark{1} and Joshua S.\
Bloom\altaffilmark{1,2,3}}

\affil{$^1$ Harvard-Smithsonian Center for Astrophysics, MC 20, 60 Garden
Street, Cambridge, MA 02138, USA}

\affil{$^2$ Harvard Society of Fellows, 78 Mount Auburn Street, Cambridge, MA
02138 USA}

\affil{$^3$ Department of Astronomy, 601 Campbell Hall, University of
California at Berkeley, Berkeley, CA 94720}

\begin{abstract} A body in Solar orbit beyond the Kuiper belt exhibits an
annual parallax that exceeds its apparent proper motion by up to many
orders of magnitude. Apparent motion of this body along the
parallactic ellipse will deflect the angular position of background
stars due to astrometric microlensing (``induced parallax'').  By
synoptically sampling the astrometric position of background stars
over the entire sky, constraints on the existence (and basic
properties) of a massive nearby body may be inferred. With a simple
simulation, we estimate the signal-to-noise ratio for detecting such a
body -- as function of mass, heliocentric distance, and ecliptic
latitude -- using the anticipated sensitivity and temporal cadences
from \gaia\ (launch 2011). A Jupiter-mass ($\mjup$) object at 2000 AU
is detectable by \gaia\ over the whole sky above $5 \sigma$, with even
stronger constraints if it lies near the ecliptic plane.  Hypotheses
for the mass ($\sim 3 \mjup$), distance ($\sim 20,000$ AU) and
location of the proposed perturber (``Planet X'') which gives rise to
long-period comets may be testable.
\end{abstract} 


\section{Introduction\label{sec:intro}}

Observations of long-period comets in the inner Solar System suggest not only a
substantial population of comets at 50,000 to 100,000 AU (the Oort Cloud;
\citealt{oort50}), but a mechanism for effectively perturbing the orbits of
these comets. Such a perturber must be massive enough to hold considerable
gravitational influence on the Oort cloud. Galactic tidal perturbations could
be the cause of a steady steam of cometary infall \citep{byl83} while close
encounters with passing stars would cause a more punctuated cascade
\citep{hills81}. Punctuated (and perhaps periodic; \citealt{hae87}) cometary
showers into the inner Solar system could also be caused by a perturber that
is bound to the Sun. Specific predictions of the mass and orbit
($\sim 0.003 M_\odot$, $d \approx 1-10 \times 10^{4}$ AU) of such a perturber
depend on whether its existence is invoked to explain temporal features in mass
extinctions on Earth (``Nemesis''; e.g.,~\citealt{dhm84},\citealt{wj84}, and
\citealt{vs93}) and/or the trajectories of anomalous streams of comets
(``Planet X''; see \citealt{murr99} and \citealt{mww99}, but see a more
cautious view from \citealt{he02}).

There are some direct constraints on the existence of any massive (planetary or
larger) perturber in the outer Solar System.  To have eluded detection by all-sky
synoptic surveys like {\it Hipparcos} and Tycho-2 \citep{hog00}, any massive
body in the outer Solar System but must be fainter than $V \sim 11$\, mag,
corresponding to absolute magnitude $M_V > 21$\,mag for $d < 0.1$ pc. 
This constraint rules out main sequence stars above the hydrogen-burning limit.

Detection of a massive perturber through reflected Solar light grows
increasingly difficult with increasing distance due to $r^{-4}$ dimming. In
reflected light, at current sensitivity limits and angular size coverages,
discoveries of objects in the Kuiper Belt at $\approx$40 AU have only recently
become routine (e.g., \citealt{btr04}). Yet even with an all-sky synoptic
survey to limiting magnitudes of  $R=24$\, mag (e.g.,
Pan-STARRS\footnotemark\footnotetext{{\tt
http://pan-starrs.ifa.hawaii.edu/project/reviews/PreCoDR/documents/
scienceproposals/sol.pdf}}), massive planets like Neptune would be undetectable
via reflected light beyond $\sim$800 AU and a 0.1 $M_\odot$ perturber with a
density of 1 g cm$^{-3}$ would be undetected with $d > 2000$\,AU.

Old and cooled degenerate stars (emitting thermally) could be faint enough to
have gone undetected. The oldest neutron star (NS) known with an apparent
thermal emission component is B0950$+$08 with $M_B \approx (20.0 \pm 0.2)$\,mag
\citep{zsk+02} ($d \approx 260~{\rm pc}$; age = 10$^{7.2}$ yr). At $d=90,000$
AU, the source would be $B \approx 13$\,mag, likely detectable with the next
generation synoptic surveys. However, with a cooling time that of the age of
the Solar System, we would expect a NS perturber to have cooled considerably,
likely to $T < 10^3$ K from $T \approx 10^5$ K (extrapolating from
\citealt{plp+04}) and so would be significantly fainter than current detection
levels.  Constraints on the existence of even colder distant planetary-mass 
objects from the
lack of detection of their thermal infrared emission with the {\it Infrared
Astronomical Satellite (IRAS)} are largely superseded by constraints from the
ephemerides of the outer planets~\citep{hogg91}.  An infrared survey with
significantly higher spatial resolution and sensitivity may provide interesting
constraints on distant objects.

Surveys that monitor distant stars with high cadence to search for
occultations by foreground objects are in principle sensitive to
objects of mass as low as $\sim 0.01~\mearth$ out to the Galactic
tidal radius of the solar system at $\sim 10^5~{\rm AU}$. However, the
probability that any one object will occult a sufficiently bright
background star to be detectable is very low.  Therefore, in order to
detect any occultation events at all, a large number of objects must be
present.  Thus such surveys can only constrain the existence of a substantial
population of objects, and will place essentially no constraints on
the existence of individual bodies in the outer solar system.

Clearly, the limits on faint massive objects in the outer Solar System must be
probed with a fundamentally different technique than through reflected, thermally
emitted, or occulted light. Here we suggest an indirect search for massive
outer Solar System bodies by observing the differential astrometric
microlensing signature that such bodies would impart on the distant stars. As
the apparent position of the lens moves on the sky, astrometric monitoring of
background sources in the vicinity of the lens (with the appropriate
sensitivity) will reveal a complex pattern of apparent motion of those
background sources.  In \S \ref{sec:ip} we introduce the microlensing formalism
in the regime of interest. Detecting the astrometric microlensing signature of
a lens requires either the background stars to move and/or the lens to move.
Nearby objects exhibit extremely large parallaxes and so the apparent position
of the lens, regardless of whether it can be detected directly in reflected
light, sweeps out a large area of influence on the sky even if the proper
motion of lens is small. Indeed, parallax dominates the apparent motion of
objects in Solar orbit beyond the Kuiper Belt. In \S \ref{sec:sn} we estimate
the detectability of a nearby massive perturber using the data from the \gaia\ mission\footnotemark\footnotetext{Launch excepted June 2011; {\tt
http://astro.estec.esa.nl/GAIA/}} using a Monte Carlo simulation. Finally, we
highlight some improvements in the detectability estimate for future work.

\section{Properties of Induced Parallax} \label{sec:ip}

Consider a distant source with parallax $\pis$ with an (angular) separation
$\bth$ from a foreground massive body with parallax $\pix$.  The foreground
body will deflect the apparent position of the centroid of the background
source relative to its unlensed position by, \be \Delta \bth =
\frac{\bu}{u^2+2}\thetae, \ee where $\bu=\bth/\thetae$ is the angular
separation of lens and source in units of the angular Einstein ring radius, \be
\thetae=(\kappa \mx \pirel)^{1/2}. \label{eqn:thetae} \ee Here
$\kappa=4G/c^2\au=8.144~{\rm mas}/\msun$, and $\pirel=\pix-\pis$ is the
relative lens-source parallax.  For the cases considered here, $\pis \ll \pix$.
 For $u\gg 1$, $|\Delta \bth| = \thetae^2/\theta$.

Due to parallax, the apparent position of the massive body will trace out an
ellipse on the sky over the course of a year.  In addition, it will have a
proper motion $\mux$ due to its intrinsic motion.  In ecliptic coordinates, the
position of the lens at time $t$, relative to time $t_0$ has components, \be
\delta\lambda_X(t)=\pix \sin(\omega[t-t_0]) + \mux(t-t_0)\cos\gamma \ee \be
\delta\beta_X(t)=-\pix\sin(\beta)\cos(\omega[t-t_0]) + \mux(t-t_0)\sin\gamma,
\label{eqn:lambet} \ee where $\beta$ is the ecliptic latitude of the object,
and $\gamma$ is the angle of the proper motion with respect to the ecliptic
plane. For orbits with zero inclination (in the plane of the ecliptic),
$\gamma=0$. We have also assumed $\pix\ll 1~{\rm rad}$.

\begin{figure} \centerline{\psfig{file=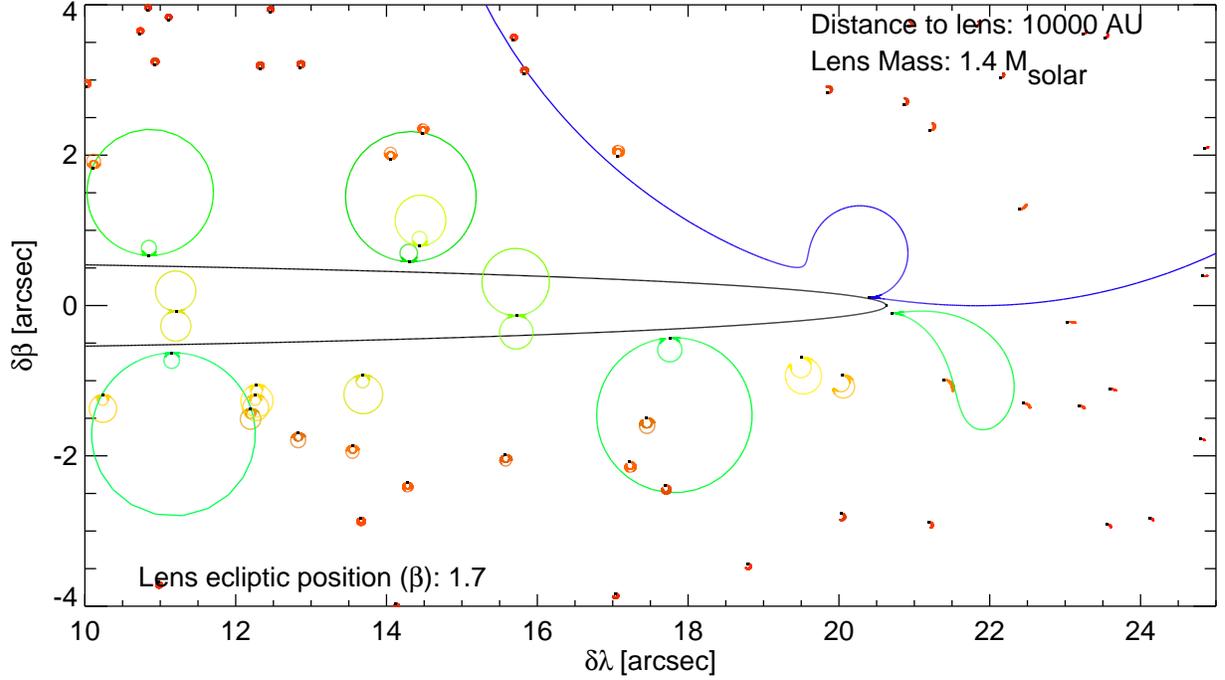,width=6.9in,angle=90}}
\caption{\label{fig:one} Illustration of the effects of a nearby lens on
background sources as the lens sweeps through its parallactic ellipse (black
solid line). The unlensed positions of the sources, assumed to be fixed and
with no parallax, are marked with small squares. The proper motion of the lens
is suppressed for illustration. Colors in the deflection tracks indicate the
relative amount of the maximum deflection (longer wavelengths correspond to
less departure from the unlensed position). This realization assumes a $M_{\rm
X}$ = 1.4 $M_{\odot}$ NS at a low ecliptic latitude ($b=1.7^{\circ}$) with a
heliocentric distance $d_{\rm X}$ = 10000 AU; the actual density of stars to
the sensitivity limit of \gaia\ is typically lower than shown here. For
other configurations of the lens, the ordinate scales as $10000$ AU/$d_{\rm X}$
and the abscissa scales as $\sin(b/1.7^\circ)/0.0297 \times (10000$ AU/$d_{\rm
X})$. The normalization of the deflection angles scales as $M_{\rm X}/1.4
M_{\odot}$, and, for the other Solar System deflector objects of interest,
would be several orders of magnitude smaller than shown here.} \end{figure}

The deflection tracks of background stars that are astrometrically microlensed
by the motion of lens parallax (hereafter ``induced parallax'') can exhibit a
variety of shapes depending on the angular position with respect to the
parallactic ellipse of the lens. Figure \ref{fig:one} shows a realization of
several tracks around a neutron star at 10,000 AU. For sources at large impact
parameter to the lens, the apparent positions over the year trace out a curved
path along a distortion angle approximately parallel with the direction of
motion of the lens at the minimum impact of the source along the parallactic
ellipse. Near the position of maximal parallactic position of the lens, these
curves resemble ``tear drop'' shapes. For impacts comparable to the semi-minor
axis of the parallactic ellipse ($\pix \sin \beta$), the deflection tracks take
the appearance of ``crescent'' shapes or a ``circle-within-circle''. This is
due to comparable deflection during the nearest impact and the distant
opposite-side impact months later; these such types of deflection paths are
obviously more common at smaller $|\beta|$. Sources interior to the parallactic
ellipse are deflected near maximally twice a year, resulting in shapes
resembling a ``figure eight''.

Although we call the deflection tracks due to parallactic motion of the lens
``induced parallax,'' the deflection tracks generally do not resemble the
traditional parallactic ellipse. First, the eccentricity of the tracks does not
generally scale with $\cos b$.  Second, the direction of motion along the
tracks is {\it retrograde} with respect to the parallactic motion of the lens. 
Moreover, unlike traditional parallax (where the date of maximum departure is
fixed by the ecliptic azimuth), the time of maximum departure from the unlensed
positions depends only on the time of minimum impact of the source to the lens.
In these ways, the source parallactic motion may be distinguished from the
effects due to induced parallax in principle. However, in practice the presence
of intrinsic source proper motion and parallax, which are typically much larger
than the signals we are concerned with here, as well as poor sampling and
signal-to-noise ratio, may cause considerable degradation of the detectability.  We
consider these issues in more detail below.

\section{Estimating the Lens Mass-Distance Sensitivity of an Astrometric Survey}
\label{sec:sn}

Figure \ref{fig:one} shows a rather dramatic effect of a nearby neutron star
upon a background field, with deflections of many background sources more than
arcseconds from unlensed positions. Since the magnitude of the deflection
tracks scales as the mass of the lens, all-sky astrometric missions could, in
principle, probe to masses significantly smaller than $M_\odot$. We now
quantify what mass/distance configurations would give rise to a detectable
signal in the presence of astrometric uncertainty and a finite number of
position samples of the background sources. Though the deflection of a single
background source may not be detectable, clearly neighboring sources will
exhibit similar, correlated deflection; therefore, the presence of a nearby
massive lens can be inferred at a statistically significant level by
aggregating a collection of statistically insignificant deflections.

Consider a massive body in solar orbit with mass $\mx$ and heliocentric
distance $\dx$.  This body will have a parallax $\pix=\au/\dx$, and a proper
motion $\mux=v_X/\dx$, where $v_X$ is its transverse velocity.  If we assume
that the body is in a circular orbit, and that $\dx\gg \au$ (so that projection
effects are small), then $v_X=v_\oplus \pix^{1/2}$.

Now consider that the body is moving in front of a background screen of source
stars with surface density $\Sigma_*$, and that series of $N$ astrometric
measurements of these stars are taken at times $t_j$.  At each time $t_j$, we
can compute the deflection due to the lens $\Delta \bth_k(t_j)=[\Delta
\theta_{\lambda,k}(t_j),\Delta \theta_{\beta,k}(t_j)]$, for each source $k$,
using the expressions presented in \S\ref{sec:ip}.  Assuming all the source
stars have the same (one-dimensional) astrometric uncertainty $\siga$, we can
estimate the total signal-to-noise ratio $\sn$ with which the deflection of the
massive body is detected as, \be (\sn)^2 = \frac{1}{2\siga^2} \sum_k \sum_j
(\Delta \theta_{\lambda,k}(t_j)-\ave{\Delta \theta_{\lambda,k}})^2+ (\Delta
\theta_{\beta,k}(t_j)-\ave{\Delta \theta_{\beta,k}})^2. \label{eqn:snone} \ee
Here $\ave{\Delta \theta_{\lambda,k}}$ and $\ave{\Delta \theta_{\lambda,k}}$
are the average deflections, i.e.\ $\ave{\Delta \theta_{\lambda,k}} \equiv
N^{-1}\sum_j \theta_{\lambda,j}$. These are the average positions of the source
determined over the course of the \gaia\ mission relative to some external
reference grid well away from the deflector.  Adopting this $\sn$ criterion for
detection is in some sense conservative, in that it only defines the
significance with which the positions of the background stars differ from the
null hypothesis of no deflections.  The effective $\sn$ will likely be
increased by fitting a model to the data which implicitly accounts for the
shape of the deflection track, as well as the correlation between neighboring
sources.  We note that, using the median deflections in \eq{eqn:snone}, rather
than the mean, increases the $\sn$ by $\sim 10\%$.

We estimate the $\sn$ using a simple Monte Carlo\footnotemark\footnotetext{Note
that it is possible, using some simplifying assumptions and by analyzing the
problem in limiting regimes, to make significant analytical progress and arrive
at simple expressions for the signal-to-noise ratio as a function of the mass
and distance to the perturber, as well as the surface density and astrometric
accuracy of the source stars. We have chosen not to present these analytic
expressions here, as there are not fully general, and so one ultimately must
resort to numerical evaluations to determine the detectability in all relevant
regimes. We note that these analytic results generally confirm the numerical
results we now present.}.  We create a random screen of stars, and simulate a
series of $N$ uniformly sampled measurements. We then calculate $\sn$ using
\eq{eqn:snone}. Under our assumptions, the $\sn$ depends on the parameters of
the lens, $\mx,\pix,\beta,t_0,\gamma$, as well as the properties of the source
stars, $\Sigma_*,\siga$.   We calculate $\sn$ for many different realizations
of the positions of the background source stars, and we also vary the input
parameters.  We find the following approximate expression for the signal-to-noise
ratio, \be
\sn \simeq \left\{ \begin{array}{ll} \frac{10\muas}{\sqrt{2}\siga}
\left(\frac{\mx}{\mearth}\right) \left(\frac{\dx}{10^3~\au}\right)^{-1}
\left(\frac{\Sigma_*}{10^{-3}~{\rm arcsec}^{-2}}\right)^{1/2}
\left(\frac{N}{40}\right)^{1/2} \left(1+\sin{\beta}\right) & {\rm if}\,\,
\Sigma_*\pi\pix^2 \ge 1 \\ 
\frac{10\muas}{\sqrt{2}\siga}
\left(\Sigma_*\pi\pix^2\right)^{1/2}
\left(\frac{\mx}{\mearth}\right) \left(\frac{\dx}{10^3~\au}\right)^{-1}
\left(\frac{\Sigma_*}{10^{-3}~{\rm arcsec}^{-2}}\right)^{1/2}
\left(\frac{N}{40}\right)^{1/2} \left(1+\sin{\beta}\right)  & {\rm if}\,\,
\Sigma_*\pi\pix^2 < 1 \end{array}\right.. \label{eqn:snscale}
\end{equation} 
The two regimes in \eq{eqn:snscale} correspond to the strong, `collisional'
regime where there is on average one star in the parallactic ellipse, and the
weak, `tidal' regime where there is typically less than one star in the
ellipse. Equation \ref{eqn:snscale} is generally accurate to considerably
better than the variance at fixed values of the parameters due to Poisson fluctuations
in the number density and location of source stars, for most parameter
combinations.  The $\sn$ can vary by a large amount due to Poisson noise
depending on the parameters, and especially so in the tidal regime for low
$\Sigma_*$. Note that, as reflected in \eq{eqn:snscale}, we find that the $\sn$
does not depend on $t_0$ or $\gamma$ to within the Poisson fluctuations.

\subsection{Application to \gaia}

In order to provide a quantitative estimate of the mass-distance
sensitivity of an astrometric survey to massive objects in the outer
solar system, we adopt parameters appropriate for the \gaia\
mission.  \gaia\ will monitor the entire sky synoptically for five
years, acquiring astrometric measurements for $O(10^9)$ stars down
to apparent magnitudes of $V\sim 20$.  For bright stars $(V\le 12)$,
\gaia\ will have a single-measurement astrometric precision limit of
$30~\muas$, whereas at $V\sim 20$, the astrometric accuracy will be
$\sim 1400~\muas$.  Typically, each star will have $100-200$
astrometric measurements, grouped in clusters of 2 to 5 measurements
each.

To proceed with our estimate,
we adopt a model of the surface density of source stars on the sky as a
function of magnitude, Galactic latitude and longitude, and a model of the
expected astrometric performance of \gaia.  This allows
us to predict the total $\sn$ with
which a object of a given mass and distance would be detected with \gaia,
at a given
location in the sky.

The expected performance of \gaia\ has and will continue to evolve, and the
final mission astrometric accuracy is therefore impossible to access currently.
For definiteness, we assume that the (one-dimensional) astrometric uncertainty
of each measurement is given by, \be\label{eqn:sig1d} \sigma_{1D}^2(V)=\left\{
\begin{array}{ll} \sigma^2_{sys}\qquad &{\rm if}\qquad V\le 12.5 \\ \sigma^2_s
10^{0.4(V-12.5)} + \sigma^2_b 10^{0.8(V-20)} \qquad &{\rm if} \qquad V > 12.5
\end{array},\right. \ee with $\sigma_{sys}=\sigma_{s}=30~\muas$ and 
$\sigma_b=1000\muas$. This form was chosen to reproduce the astrometric
accuracies from Table 1 of \citet{be02}. \gaia\ will not make
astrometric measurements uniformly across the sky; certain ecliptic
latitudes will be sampled a larger number of times than others.  We
assume that the number of samples as a function of ecliptic latitude
$\beta$ is given by, \be N_{samp} =
100+300\exp\left[-\left(\left|\frac{|\beta|-35^\circ}{10^\circ}\right|\right)^{
1/2}\right]. \label{eqn:nsamp} \ee This form was chosen to
qualitatively reproduce Figure 5 of \citet{be02}. We assume that the
samples are clustered into groups of $n_c$ points, and so the
effective number of points is $N=N_{samp}/n_c$, and the effective
astrometric accuracy of each point is
$\siga=\sigma_{1D}(V)/\sqrt{n_c}$.  This assumes that the
single-measurement errors can be reduced by root-$n$ averaging.  This
may not be the case: the measurement errors in any given cluster may
be correlated, or there may exist systematic errors that are not
reducible.  Since it is difficult to anticipate the behavior of the
astrometric errors in advance, we will adopt the assumption of
root-$n$ averaging for simplicity.  We adopt $n_c=5$ \citep{be02}.
For other values of $n_c$, the $\sn$ for any given star, as well as
the integrated $\sn$, will scale as $\sqrt{n_c/5}$.

We determine the surface density of source stars as a function of position and
magnitude using a simple model for the Galaxy.  For the density distribution of
sources, we adopt the double-exponential disk plus barred bulge model of
\citet{hg95,hg03}.  We assume that the dust column is independent of
Galactocentric radius and has an exponential distribution in height above the
plane with a scale height of $120~{\rm pc}$.  We normalize the midplane column
density so that the $V$-band extinction is $A_V=1~{\rm mag}~(D_s/{\rm kpc})$,
where $D_s$ is the distance to the source. We also will show results assuming
the dust model of \citet{be02}, which is similar to ours for $\beta \ga
20^\circ$, but differs in detail for latitudes closer to the plane.  Finally,
we assume a $V$-band luminosity function that is independent position and is
equal to the Bahcall-Soneira \citep{bs80} luminosity function for $M_V\le 10$,
and is constant for $10\le M_V \le 20$.  

The surface density of stars down
to $V=20$ in our model ranges
from $\sim 10^{-5}~{\rm arcsec}^{-2}$ near the Galactic poles, to 
$\sim 10^{-3}~{\rm arcsec}^{-2}$ near the Galactic anticenter, to 
a maximum of $\sim 0.1~{\rm arcsec}^{-2}$ within a few degrees of the Galactic
center.  Therefore, regions of the sky near the Galactic plane and especially the Galactic
center will have greater sensitivity to lower mass and/or more distant perturbers
for fixed $\sn$. The total number of stars in the sky
with $10\le V\le 20$ in this model is $1.3\times 10^9$ for our standard dust
extinction model, and $1.0\times 10^9$ for the \citet{be02} dust model.
Thus the average surface density is $\sim 10^{-3}~{\rm arcsec}^{-2}$.

\begin{figure} \plotone{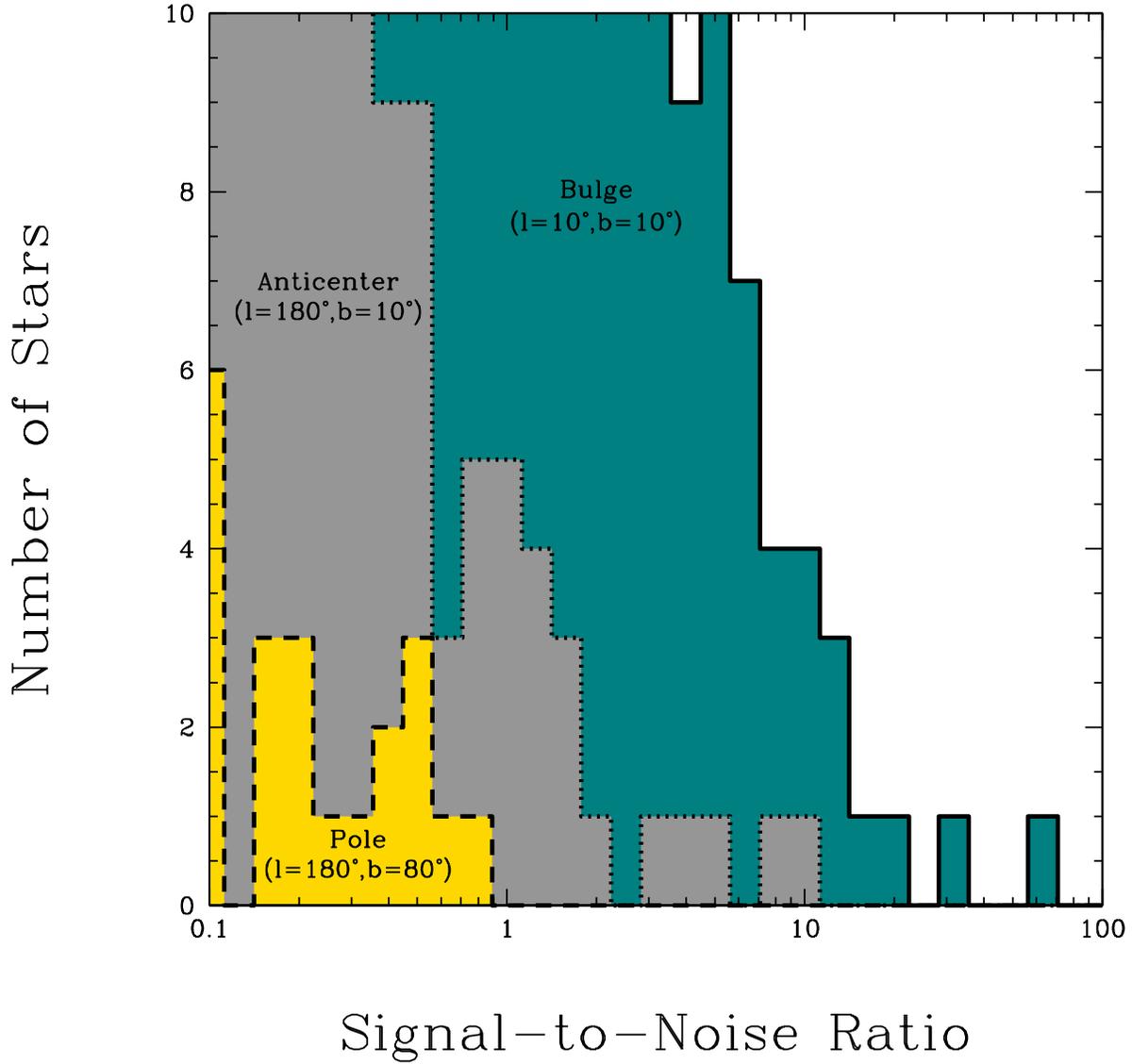} \caption{\label{fig:two} 
Distribution of signal-to-noise ratios ($\sn$) for 
detecting a
massive object of mass $M=3000\mearth\simeq 10\mjup$ and distance
$\dx=10^4~\au$ with \gaia, for 
various
locations on the sky.  The histograms show the number of stars with $10\le V\le 20$ as 
a function of the $\sn$, for three different locations for the massive
object with very different background source
densities: near the Galactic bulge (solid, $l=10^\circ,b=10^\circ$, $\Sigma_*\sim 10^{-2}$),
near the Galactic anticenter (dotted, $l=180^\circ,b=10^\circ$, $\Sigma_*\simeq 10^{-3}$), and
near the north Galactic pole (dashed, $l=180^\circ,b=80^\circ$, $\Sigma_*\simeq 10^{-5}$).  
For these three locations, an object with $M=3000\mearth\simeq 10\mjup$ and 
$\dx=10^4~\au$ would be detected at $\sntot=94.4$ (bulge), 16.5 (anticenter), and 1.4 (pole).
}\end{figure}

Figure \ref{fig:two} shows the distribution of $\sn$ for
an object with $M=3000\mearth\sim 10~\mjup$
and $D=10^4~\au$ located in three different locations on the sky:
near the Galactic bulge, anticenter and north Galactic pole.  The source
densities in these three locations vary considerably, from $\Sigma_*
\sim 10^{-2}~{\rm arcsec}^{-2}$ near the Galactic bulge to $10^{-5}~{\rm arcsec}^{-2}$ near the pole.  The
shape of the distribution of $\sn$ depends on the location on the sky,
through the distribution of source densities as a function of
magnitude (and so astrometric accuracy).  For the locations near the
Galactic plane with high source densities, the distribution of $\sn$
has a tail toward higher values, and so the total $\sn$ is generally
dominated by one or two stars.  For the location near the Galactic
pole, a larger number of stars contribute significantly to the total
$\sn$.  The total $\sn$ (integrated over $V$-magnitude from $V=10$ to $V=20$)
for these three locations are $\sntot=94.4$
(bulge), 16.5 (anticenter), and 1.4 (pole).

\begin{figure} \plotone{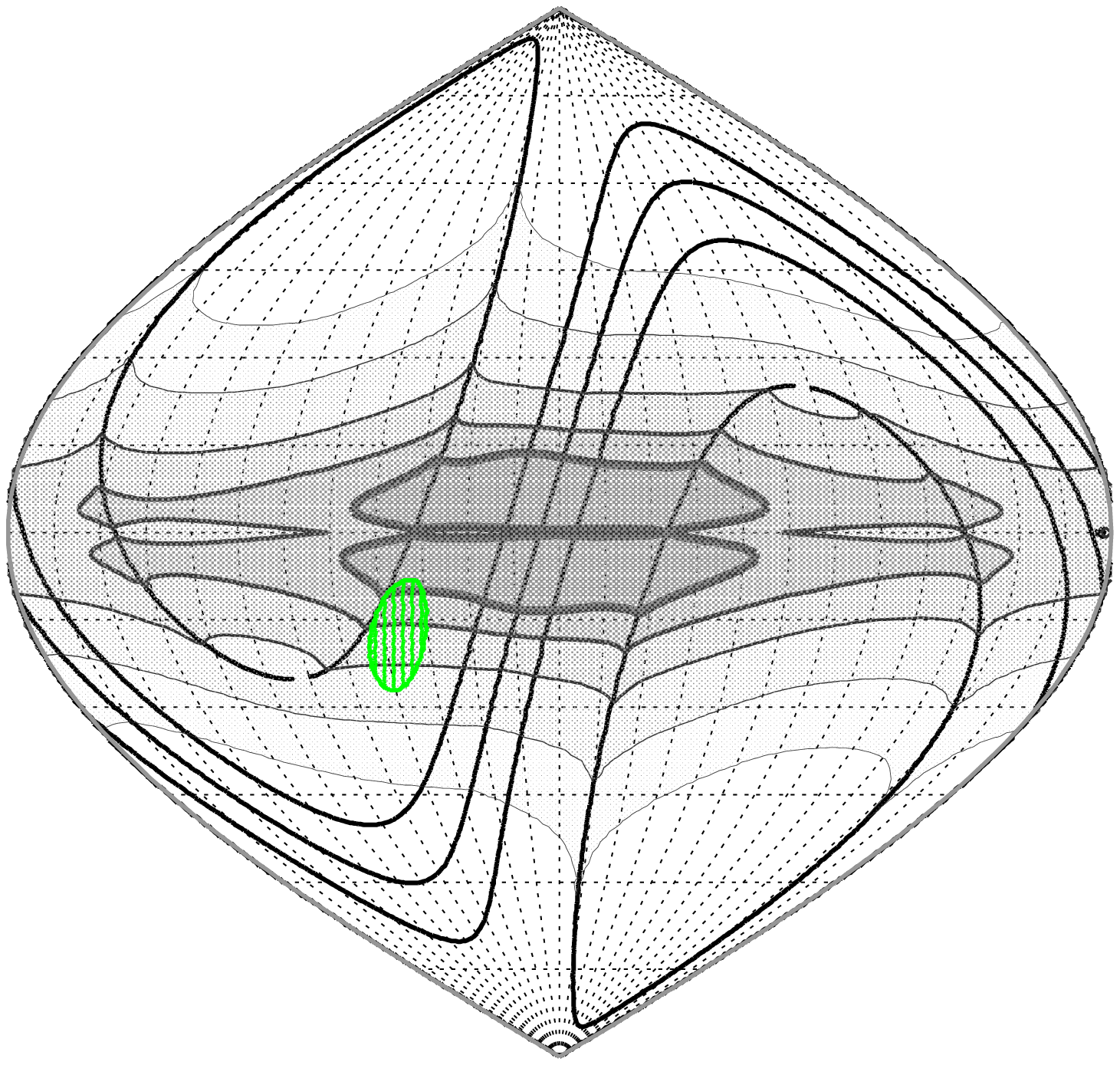} \caption{\label{fig:three} 
All sky map in Galactic coordinates of the $\sn$ for detecting a
massive object of mass $M=3000\mearth\simeq 10\mjup$ and distance
$\dx=10^4~\au$ with \gaia.  The dotted lines show
lines of constant Galactic latitude and longitude at $15^\circ$ intervals.  
The Galactic center is located at
the center of the figure.  Contours of constant $\sn$ are in grey, at
levels of $\sn=3,5,10,20,40$ (lighter to heavier).  We also show lines
of constant ecliptic latitude for $\beta=\pm 35^\circ, \pm 10^\circ,
0$ (solid black lines).  The oval shaded region brackets the uncertainty in the inferred
position of ``Planet X'' from the clustering of cometary aphelion
distances \citep{murr99}.
}\end{figure}

Figure \ref{fig:three} shows contours of constant $\sntot$ for a
object with $M=3000\mearth\sim 10~\mjup$ and $D=10^4~\au$.  The
distribution of $\sntot$ on the sky is highly non-uniform: objects of
a given $M$ and $\dx$ located toward certain regions of the sky will
be detected with higher $\sntot$ than if they are located in other
regions. The $\sntot$ is primarily driven by the surface density of
stars, and therefore regions of the sky near the Galactic plane and
especially the Galactic center are preferred.  However, it is also the
case that the number of samples $N_{samp}$ depends on ecliptic
latitude, such that stars with ecliptic latitude $\sim \pm 35\deg$
will have several times more astrometric measurements than stars near
the ecliptic poles.  Therefore locations near ecliptic
latitudes of $\pm 35\deg$ will also have higher $\sntot$ for fixed perturber
mass and distance.

\begin{figure} \plotone{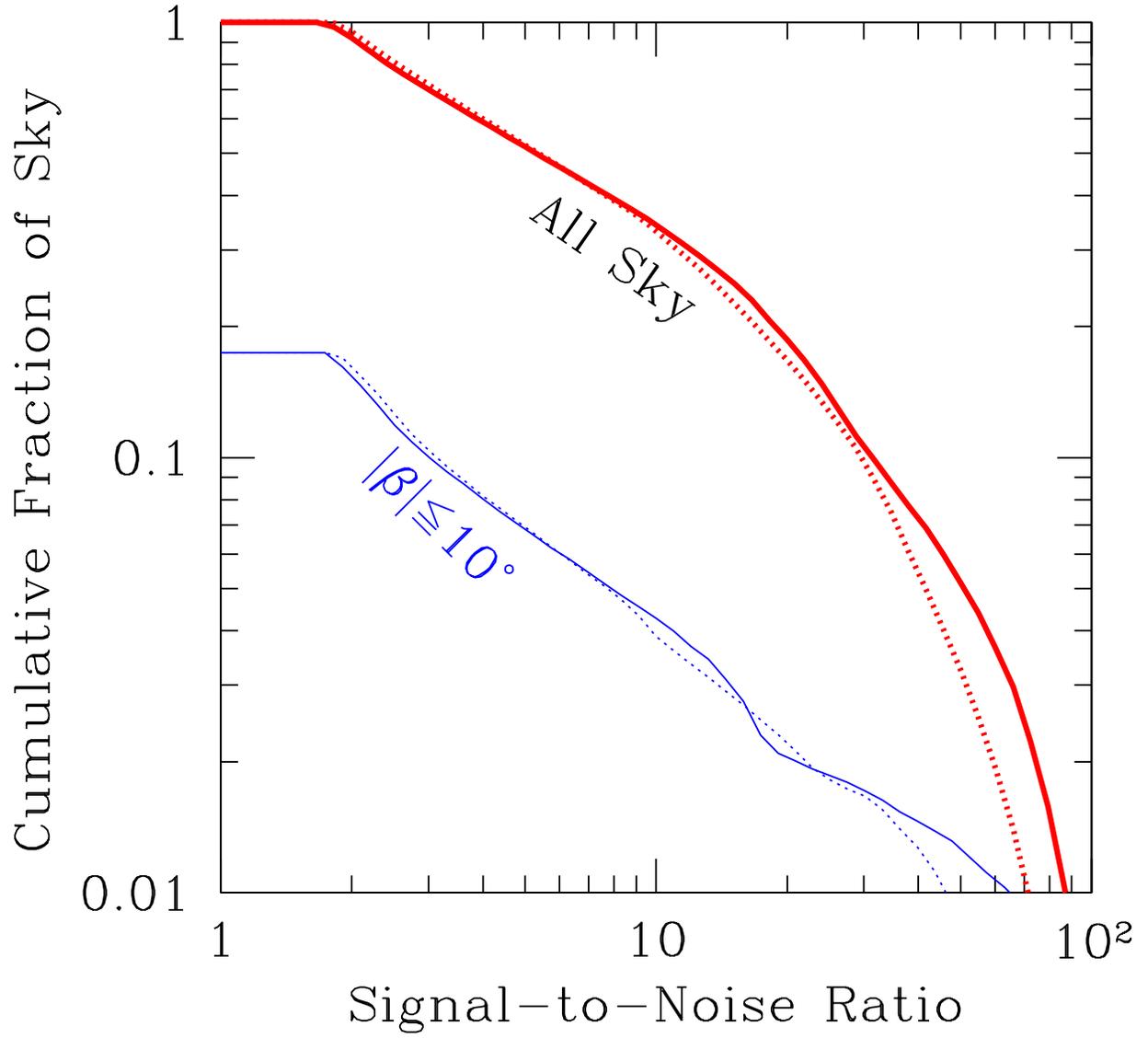} \caption{\label{fig:four} Fraction of the sky
over which an object of mass $M=3000\mearth\simeq 10\mjup$ and distance
$\dx=10^4~\au$ would be detected with $\sn$ greater than a given value.  The
solid curves show the fraction assuming our fiducial model for the dust
distribution, whereas the dotted lines shows the alternative model of
\citet{be02}. Upper curves are for the entire sky, the bottom curves are for
ecliptic latitudes $|\beta|<10^\circ$. } \end{figure}

\begin{figure} \plotone{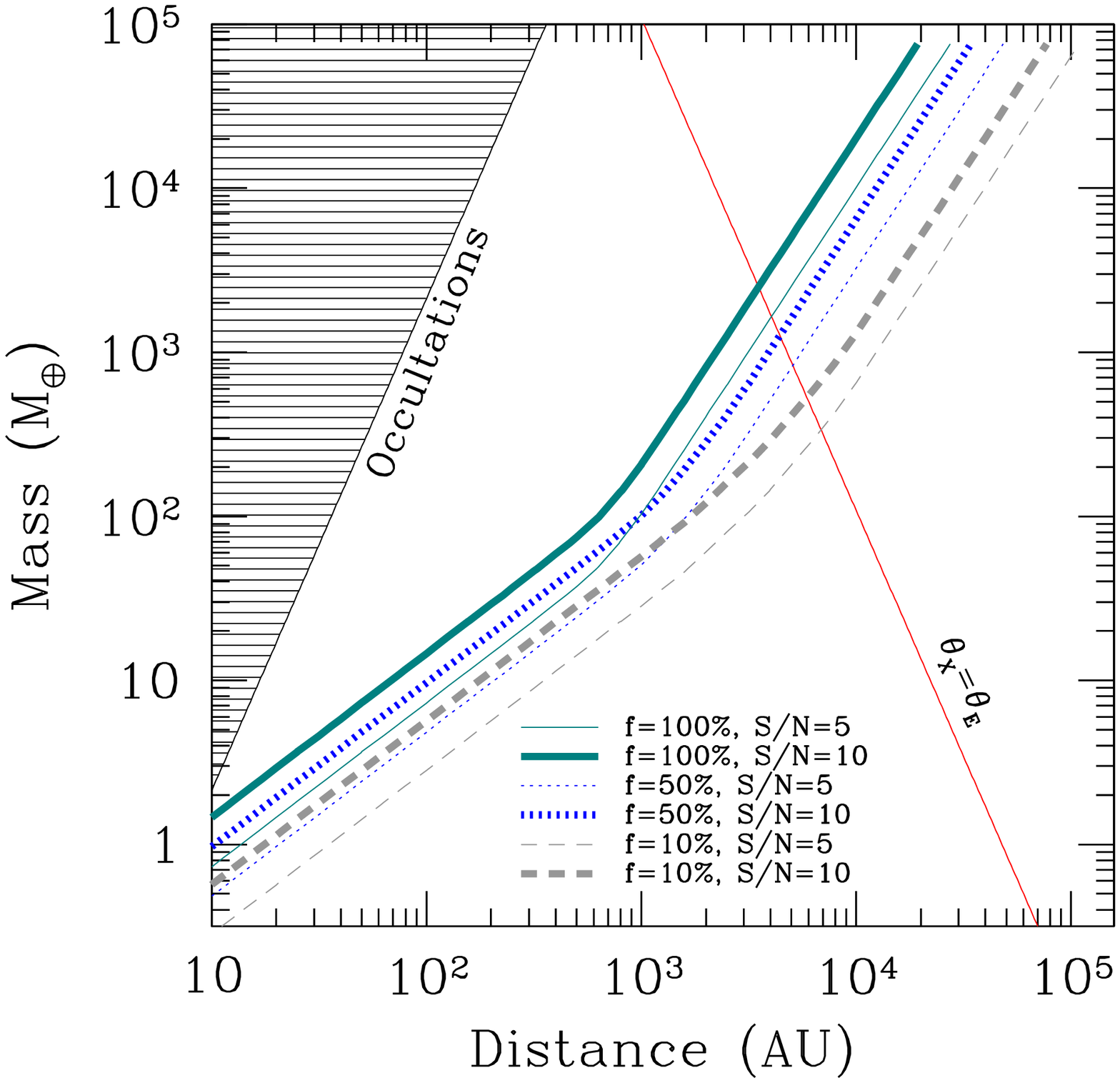} \caption{\label{fig:five} The lines show 
constraints on the mass and distance of an object that can be detected by
\gaia\ at various $\sn$ thresholds over various fractions of the sky.  The red
line shows where the angular size $\theta_X$ of an object with density of
$1~{\rm g~cm^{-3}}$ is equal to its Einstein ring radius $\thetae$; objects to
the left of this line have $\theta_X\ge \thetae$.   Objects with parameters in
the shaded region will occult at least one measurement of at least one
background source, assuming a typical background source density of $2\times
10^{-3}~{\rm arcsec}^{-2}$.  
}
\end{figure}

Figure \ref{fig:four}
shows the fraction of the sky enclosed by contours of a given $\sntot$, i.e.\
the fraction of the sky over which an object of mass $M=3000\mearth$ and
distance $\dx=10^4~\au$ would be detected with $\sn$ greater than a given
value.  We determine the fraction of sky above a given $\sntot$ for a range of
masses and distances. Objects with mass greater than a minimum mass \be M_{min}
\simeq (290,490,750)M_\oplus \left\{ \begin{array}{ll}
\left(\frac{\dx}{10^4~\au}\right) \left[\frac{(\sn)_{\rm th}}{5}\right] & {\rm
if}\, \dx\le\db \\ \left(\frac{\dx}{\db}\right)
\left(\frac{\dx}{10^4~\au}\right) \left[\frac{(\sn)_{\rm th}}{5}\right] & {\rm
if}\, \dx> \db \end{array}\right\},\, {\rm for}~\fsky=(10\%,50\%,100\%).
\label{eqn:mscale} \end{equation} can be detected with $\sn \ge (\sn)_{\rm
th}$,  where $\fsky$ is the fraction of the sky. Here $\db$ is the `break
distance', and has values of $\db=(4470,1550,780)\au$ for
$\fsky=(10\%,50\%,100\%)$. These limits are shown in Figure~\ref{fig:five}.

Figure \ref{fig:four} also shows the fraction of the sky within $10^\circ$ of
the ecliptic plane enclosed by contours of a given $\sntot$. Since the ecliptic
plane fortuitously passes near the Galactic bulge, the slope for this curve is
shallower than that for the entire sky, resulting in a relatively larger
fraction of the area for which a high $\sn$ detections are possible.

There are several obvious limitations of our calculations.  One is that we have
neglected the motion of background stars due to parallax. To the extent that
these motions correlate with the microlensing signal, they will tend to degrade
the signal-to-noise ratio, by effectively allowing one to partially `fit out' the
anomalous excursions.  Motions of stars in binaries could also confound a clean
measurement of induced parallax. Also, since we adopted the simple scaling
relation in \eq{eqn:snscale} when integrating over the magnitude distribution
of source stars, we have neglected the effect of the Poisson fluctuations of
the surface density and location of stars on the total signal-to-noise ratio. 
To provide a rough
estimate of the magnitude of these effects, we have performed a few simulations
where we determine the signal-to-noise ratio for stars of a given magnitude
directly from the Monte Carlo simulation (which {\it per force} includes
Poisson fluctuations), while explicitly fitting for the parallax of the source
stars.  Since these calculations are extremely time intensive, we have not
performed a comprehensive exploration, but rather checked only a few cases. 
For these few cases, we find that fitting for the parallax of the source
does indeed reduce $\sntot$, but by a relatively small factor, $\sim 10\%$.  
On the other hand, we find that the effect of Poisson fluctuations causes us to
{\it underestimate} $\sntot$, by as much as $\sim 75\%$.

\section{Discussion and Conclusions}

We have shown that a substantial, as yet unexplored, region of mass-distance
parameter space of nearby massive bodies will be accessible with the current
incarnation of the datastream from the \gaia\ experiment. We have focused
on the effect of ``induced parallax'' caused only by the parallax of the lens
as it sweeps through the parallactic ellipse. Based on our albeit simplistic
simulation, the search for massive bodies in the outer Solar System by the
observation of induced parallax has a reasonable chance of uncovering the
proposed perturber of cometary orbits in the Oort cloud (Figs.~\ref{fig:four}
and \ref{fig:five}). In particular, we believe that the non-detection of a massive
body in the \gaia\ dataset using the proposed technique would relegate the proposed
mass-distances of Planet X to a significantly smaller
parameter space then the currently allowed space\footnotemark\footnotetext{It
is noteworthy that \citet{he02} also appeal to \gaia\ for constraining the
existence of Planet X, but by making use of ephemeris data of $\sim$1000
long-period comets that would be discovered by \gaia\ relatively uniformly
over the sky.}.

\begin{figure} 
\epsscale{0.8}
\plotone{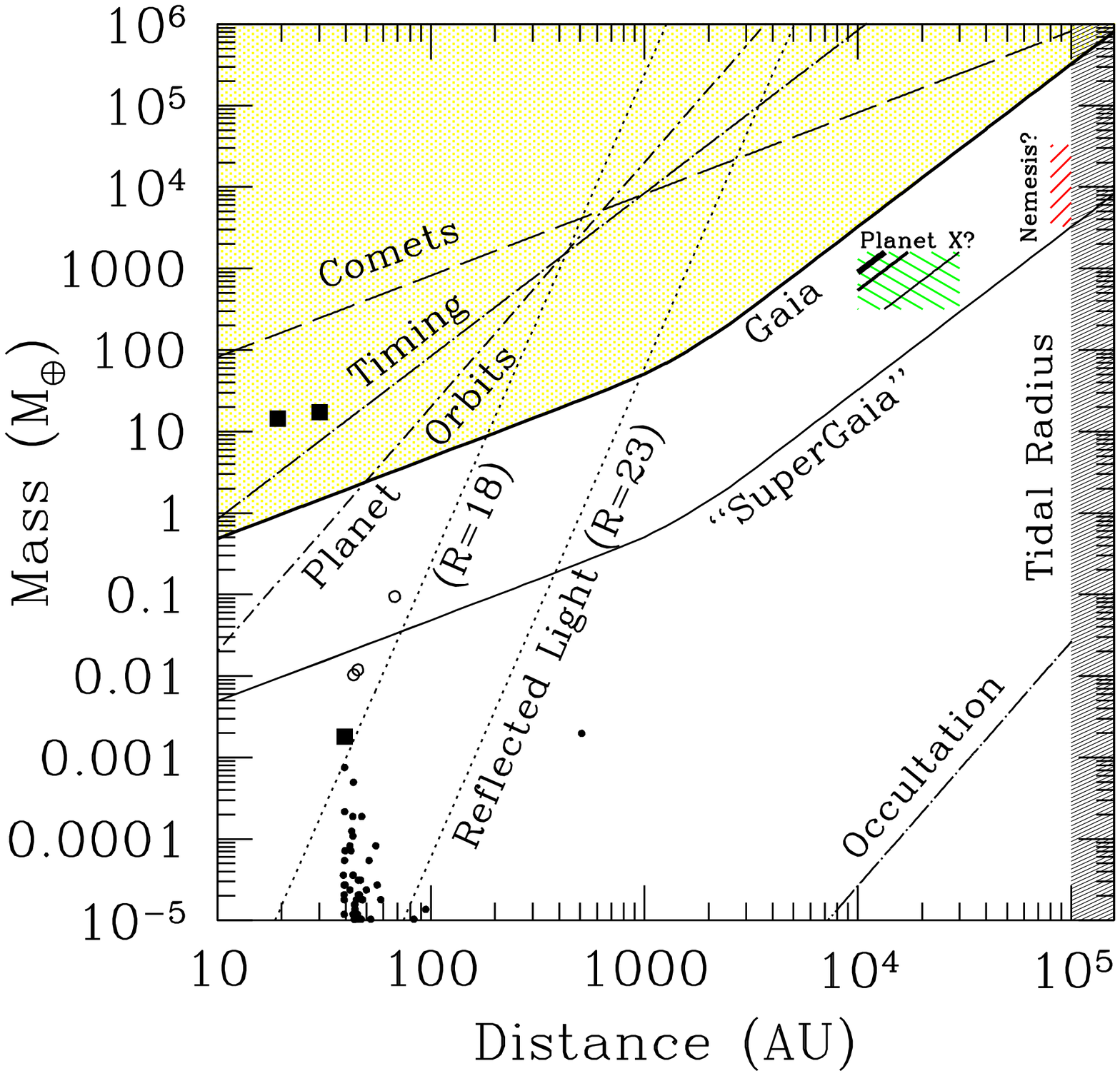} \caption{\label{fig:six} All-sky constraints
on distant massive solar-system objects using various methods.  Masses and
distances to the upper left of the lines are currently excluded by these
methods (`Comets,' `Planet Orbits,' and `Timing'), 
or can be excluded in the future (`Reflected Light,' `Gaia,' and `Occultation'). 
Limits marked `Comets' and `Planet
Orbits' were taken from \citet{hogg91}; the
limit marked `Timing' is from \citet{zakamska05}; the `Occultation' limit is derived
following \citet{gaud04}.  The line marked `Gaia' shows the mass and distance of
an object that can be detected by \gaia\ at $\sn=5$ over 50\% of the sky. The
line marked ``SuperGaia'' shows the same limit for a hypothetical experiment
with astrometric sensitivity that is two orders of magnitude better than \gaia\
with the same limiting magnitude.  The small points show the masses and
distances of known Kuiper Belt Objects (KBOs) and minor planets, where the masses were
derived from their absolute magnitude assuming an albedo of $4\%$ and a density
of $1~{\rm g~cm^{-3}}$.  The open circles show the masses of the three recently-discovered
bright KBOs 2003 EL$_{61}$, 2003 UB$_{313}$ and 2005 FY$_9$ (see \citealt{brown05} and references therein), under the same assumptions.   The large squares show the masses and distances of
Uranus, Neptune, and Pluto.  The shaded region at the extreme right shows the
tidal radius of the solar system, $\sim 10^5~{\rm AU}$. The shaded 
rectangles show the range of masses and distances inferred for `Planet X'
and `Nemesis.'  Since the proposed ecliptic longitude and
latitude of `Planet X' \citep{murr99} is near the Galactic plane, the 
limiting masses
probed in that region are smaller than over the sky as a whole: the short lines
passing through the shaded region labeled `Planet X' show the lower limits on the mass and
distance of an object that yields $\sn=1,3,5$ (lighter to heavier) at the
\citet{murr99} position.
}
\end{figure}

\citet{murr99} made specific predictions for the current position of ``Planet
X'' on the sky, based on the clustering of cometary aphelion distances.  Since
the $\sn$ map of the sky is not uniform, it is interesting to ask with what
$\sn$ one would expect to detect ``Planet X'' with the allowed mass and
distances, at its expected position.  Figure \ref{fig:three} shows the positional
error ellipse from \citet{murr99}.  The expected $\sn$ for $\mx=10^3\mearth$
and $\dx=10^4\au$ ranges from $\sntot\simeq  3$ to $\sntot \simeq 12$.  The
mass/distance limit for thresholds of $(\sn)_{th}=1,3$, and $5$ in this error
ellipse is shown in Figure \ref{fig:six}; roughly 25\% the allowed parameter
space could be excluded at $3\sigma$ with a non-detection.

Hypothesis for the mass ($\sim 0.03\msun$) and distance ($\sim
10^5~\au$) of Nemesis will likely be difficult to test with \gaia\ (see
Fig.~\ref{fig:six}), due primarily to the large distance and thus
small size of the parallactic ellipse.  However, specific predictions
for the current position of Nemesis might be testable using a targeted
astrometric satellite with higher astrometric precision than \gaia,
such as the {\it Space Interferometry Mission (SIM)}.  Of course,
constraints on smaller-mass objects at any distance could be obtained with
with an all-sky synoptic experiment that has improved astrometric
accuracy but with a similar limiting magnitude (``SuperGaia'',
Fig.~\ref{fig:six}) or by probing more stars to fainter levels with
\gaia-like astrometric accuracies.

Should a significant detection be made, what can be learned about the lens?  In
principle, the astrometric data alone provide an estimate of the mass,
position, distance, and proper motion of the lens for high-$\sn$ detections of
induced parallax for stars very near to the parallactic ellipse.  Orbit
determination will generally be difficult, unless there is a significant
acceleration over the five year mission lifetime; this is only expected for
relatively nearby lenses.  For more modest $\sn$ detections, or detections in
the tidal regime where the source stars are quite distant from the parallactic
ellipse, the information will be seriously degraded, and degeneracies between
the mass, distance, and angular separation from the lens arise.  In the extreme
case where only one distant star is significantly perturbed, the detection may
yield very little information about the lens.  Exploration of the information
that can be extracted from these various classes of detections is beyond the
scope of this paper, but is an interesting topic for future study.

Further follow-up of potential candidates may be possible with a variety of
methods.  Astrometric follow-up of individual background sources may be
possible with {\it SIM} with higher
astrometric precision and cadence than possible with \gaia; such measurements
may improve on the determination of the lens parameters.  Direct detection of
the reflected light from some candidates may be possible with ultra-deep
imaging using very large aperture, next generation, ground-based,
optical/near-infrared telescopes such as the {\it Giant Magellan Telescope
(GMT)}, the {\it Thirty Meter Telescope (TMT)}, or the {\it Overwhelmingly
Large Telescope (OWL)}.  Finally, the {\it James Webb Space Telescope (JWST)}
should have the sensitivity to detect the thermal emission from essentially all
objects detected astrometrically by \gaia.

A similar astrometric microlensing search with \gaia\ for massive stellar
remnants in the Solar neighborhood ($d \sim 150$\,pc) was proposed by
\citet{be02} but with several important differences compared to the present
work. First, we considered the detectability of an object significantly closer
to Earth so that the lens parallax is $\sim10^{5-7}$ larger than the typical
source parallaxes whereas that difference is only $10^{1-2}$ for Solar
neighborhood lenses. We also focused on Solar System lenses in Solar orbit
where the parallax motion dominates proper motion; the motion of Solar
neighborhood objects are dominated by proper motion. Both these different
regimes result in significantly different microlensing tracks of a single
background star (compare our Fig.~\ref{fig:one} with Fig.~1 of \citealt{be02}).
Second, we focus on detection of objects with a planet-scale mass whereas the
analysis technique of \citealt{be02} is optimized to constrain the mass
function of stellar-mass objects in the Solar neighborhood (see, e.g., Fig.\
\ref{fig:three}) with $M > 0.1 M_\odot$. Last, and conceptually the most
distinct, we consider the detectability of a single massive object using the
aggregate induced parallax signatures of thousands of stars whereas
\citeauthor{be02} focused on constraining the properties of a large population
of faint stellar-mass objects, where the mass of each object is inferred using
the astrometric microlensing ``event'' a single background source. Ultimately,
though, both analyzes make use of the same datastream and act toward
complimentary goals.

We have assumed that our lenses are point-like, and so have ignored the effects
of occultation of the background sources by the lens.  If the angular size of
the lens $\theta_X$ is an appreciable fraction of its angular Einstein ring
radius $\thetae$, then both occultation and lensing effects can potentially be
important \citep{agol02,taka03}. In Figure \ref{fig:five}, we show the locus of
mass and distance where $\theta_X=\thetae$.  Objects with $M_X\la 10^3~\mearth$
will have angular sizes that are larger than their Einstein ring radii provided
they are closer than $\sim 4000~\au$; for such objects, complete occultations
are possible.  However, an occultation will obviously only occur if a
background source happens to be located within an angular radius of the lens
when a measurement is taken. This condition is met when the number of
measurements satisfies $N\Sigma_* \pi \theta_X^2\sim 1$.  Figure \ref{fig:five}
shows the region of parameter space for which at least one measurement will be
occulted by the lens, for typical background source densities of
$\Sigma_*=2\times 10^{-3}~{\rm arcsec}^{-2}$. Clearly, for most lenses,
occultation effects are negligible.

In our simulation, we assumed the perturber is in a circular orbit around the
Sun.  However, we found that our results are essentially independent of the
proper motion of the lens.  Furthermore, realistic motions along the
line-of-sight are unlikely to alter our signal-to-noise ratio estimates
substantially for the distances considered herein.  Therefore, the assumption
that the lens is on a circular orbit or indeed even bound to the Sun is
immaterial to our conclusions.

As we have discussed, an obvious shortcoming of our estimation is that we have
neglected the motion of background stars due to parallax, proper motion, and
orbits. These motions will tend to degrade the signal-to-noise ratio, effectively
introducing more free parameters to help explain away anomalous excursions. 
Still our preliminary calculations show that source parallax is not likely to
degrade the $\sn$ substantially, however these calculations were admittedly
limited.  We hope to perform a more comprehensive study to quantify the effect
of a realistic background screen in future work.

Our simplistic simulation for S/N estimation also neglects another feature of
data that could be exploited to {\it improve} the S/N.  Any nearby foreground
massive source will lens multiple source background stars differently in the
course of a 5 year mission. Moreover, neighboring background sources will be
lensed similarly. So the expectation of correlated deflection paths (which are
fixed for a given lens mass, distance, and proper motion) could be used to
create a ``matched filter'' for improving the sensitivity of detecting a nearby
massive lens. Though computationally very expensive, one can envision applying
such a filter to the \gaia\ dataset for all possible nearby lens masses at
all possible distances and positions on sky to search for a signal. Aside from
the need to simultaneously constrain the parallax, proper motion, and orbital
parameters of all background sources, the matched filter search may also need
to search for a possible changing parallax of the lens over the mission
lifetime: a massive object passing nearby that  is unbound to the Sun with $|v|
\approx 30$ km s$^{-1}$ would travel $\approx$30 AU over 5 yr, with some of
this motion in the radial direction from the Sun.

Finally, the choice of the appropriate $\sn$ threshold for a robust detection
deserves some discussion.  Here one must not only consider the astrometric
noise properties of the sources, but also the total number of independent
trials performed in searching the data with a matched filter.  This latter
factor can be quite crucial in the current context, given the fact that one is
performing a blind search over the entire sky with $O(10^9)$ source stars, with
many independent filters corresponding to varying lens locations, distances,
masses, and proper motions.

While a high signal-to-noise ratio measurement of the entire induced parallax of
single star will yield the lens mass, sky position, proper motion, and
distance, the likelihood of such a special configuration is rare.  Instead,
each of these lens events will contribute individually to constraints on the
lens properties at different times, leading to the possibility of improving the
signal-to-noise ratio of the lens properties through the matched filter. Another
utility of global astrometric filtering of the \gaia\ data is that the
masses and ephemerides of known Solar System objects might be determined {\it a
priori}, based solely on measurements of the astrometric microlensed
background; whether the masses determined thusly will be more precise than
measured by other means remains to the be seen.

\acknowledgments BSG supported by a Menzel Fellowship from the Harvard College
Observatory. JSB was partially supported by a grant from the
Harvard-Smithsonian Center for Astrophysics. We would like to thank Avi Loeb
for comments on the manuscript, and the anonymous
referee for a prompt and helpful report.  We would like to extend special thanks to Andy
Gould for his detailed and comprehensive comments and insightful discussions
which led to a much improved paper.  Lastly, JSB thanks Eugene Chiang and Josh
Eisner for their enthusiasm during the early stages of this work.

\end{document}